\documentclass[12pt]{iopart}

\usepackage{color}
\usepackage[english]{babel}
\usepackage{graphicx,amsmath,amssymb,ulem}

\newcommand{\ket}[1]{\left|#1\right>} 
\newcommand{\bra}[1]{\left<#1\right|} 
\newcommand{\ee}{\mathrm{e}} 
\newcommand{\ii}{\mathrm{i}} 

\begin{document}
\title[Interference of macroscopic beams on a beam splitter]{Interference of macroscopic beams on a beam splitter: phase uncertainty converted into photon-number uncertainty}

\author{K. Yu. Spasibko$^{1,2,\footnotemark[1]}$, F. T\"oppel$^{2,3,4,\footnotemark[1]}$, T. Sh. Iskhakov$^{2}$, M. Stobi\'nska$^{5,6}$, M. V. Chekhova$^{2,3,1}$, and G. Leuchs$^{2,3}$}
\footnotetext[1]{These two authors contributed equally}
\address{$^1$Faculty of Physics, M. V. Lomonosov Moscow State University, 119991 Moscow, Russia\\
$^2$Max-Planck-Institute for the Science of Light, G\"unther-Sharowsky-Stra{\ss}e 1/Bldg. 24, 91058 Erlangen, Germany\\
$^3$Institute for Optics, Information and Photonics, Universit\"at Erlangen-N\"urnberg, Staudtstra{\ss}e 7/B2, 91058 Erlangen, Germany\\
$^4$Erlangen Graduate School in Advanced Optical Technologies (SAOT), Paul-Gordan-Stra{\ss}e 6, 91052  Erlangen, Germany\\
$^5$Institute of Theoretical Physics and Astrophysics, University of Gda\'nsk, ul.~Wita Stwosza 57, 80-952 Gda\'nsk, Poland\\
$^6$Institute of Physics, Polish Academy of Sciences, Al.~Lotnik\'ow 32/46, 02-668 Warsaw, Poland}

\ead{kirill.spasibko@mpl.mpg.de}

\begin{abstract}
Squeezed-vacuum twin beams, commonly generated through parametric down-conversion, are known to have perfect photon-number correlations. According to the Heisenberg principle, this is accompanied by a huge uncertainty in their relative phase. By overlapping bright twin beams on a beam splitter, we convert phase fluctuations into photon-number fluctuations and observe this uncertainty as a typical `U-shape' of the output photon-number distribution. This effect, although reported for atomic ensembles and giving hope for phase super-resolution, has been never observed for light beams. The shape of the normalized photon-number difference distribution is similar to the one that would be observed for high-order Fock states. It can be also mimicked by classical beams with artificially mixed phase, but without any perspective for phase super-resolution. The probability distribution at the beam splitter output can be used for filtering macroscopic superpositions at the input.

\end{abstract}

\pacs{42.50.-p, 42.50.Ar, 42.50.Dv, 42.25.Hz}
\normalsize

\maketitle

\def\thesection{\arabic{section}}

\vspace{5mm}

\maketitle

\section{Introduction}


Interference of nonclassical light~\cite{Marek2012} is a fascinating phenomenon, playing the central role in quantum optics and quantum information research. In particular, important effects are observed by overlapping light beams on a balanced beam splitter, which transforms amplitude fluctuations into phase fluctuations and vice versa~\cite{Klyshko,Pfister}. The most widely used is the Hong-Ou-Mandel (HOM) effect, which occurs when two indistinguishable photons arrive at the inputs: they both exit from the same output port~\cite{Mandel,Alley}. As a result, the probability for two detectors each located at one output port of the beam splitter to click in coincidence is zero.

A direct generalization of the HOM effect for highly populated photon-number states was analyzed theoretically in Ref.~\cite{Campos}. It was shown that for two identical Fock states entering a balanced beam splitter, only even photon numbers are observed at each output port. Moreover, the photon-number probability distribution has a characteristic concave shape (`U-shape'), with the width scaling as the mean photon number. This property is useful for quantum filtering and engineering of macroscopic quantum superpositions of light~\cite{filter} as well as loophole-free Bell inequality tests~\cite{Stobinska2011,Stobinska2013,Stobinska2013a}.

The reason why the U-shape is observed in the output photon-number distribution is that for the input identical Fock states, the photon-number difference has zero uncertainty. Since the phase difference between the input fields is Heisenberg conjugate to the amplitude difference~\cite{Holland}, it has huge uncertainty, which is transformed by the beam splitter into the uncertainty of the photon-number difference at the output~\cite{Pfister}. A similar shape should be observed for the case of twin beams at the input, as their photon-number difference is very well
defined~\cite{Pfister}. In fact, this behavior has been predicted~\cite{Kasevich} and discovered~\cite{BEC} for Bose-Einstein condensates with pair correlations. As, in its turn, the huge photon-number uncertainty can mean, for a sufficiently pure state, reduced phase uncertainty after the beam splitter, the U-shape was considered in the literature as an indication for possible phase super-resolution~\cite{Holland,Pfister,Kasevich,BEC}. Surprisingly, it has been never observed for light twin beams.


In this paper, we demonstrate this effect for twin-beam bright squeezed vacuum, produced in a traveling-wave parametric amplifier. Although the total photon number in each of the twin beams is uncertain, there is perfect (ideally, up to a single photon) photon-number correlation between the two beams.
Our results confirm the predicted U-shape of the photon-number probability distribution at the beam splitter outputs. Due to the losses and the absence of single-photon resolution we are not able to show the odd/even discrete structure of the distribution. Measurements with different numbers of modes show that the U-shape can be only observed for single-mode or nearly single-mode beams. Finally, we show that by artificially varying the phase of a classical source one can obtain a similar distribution, which, however, does not indicate any initial photon-number correlations.

This paper is organized as follows. In Section \ref{1} we briefly review the properties of the twin-beam bright squeezed vacuum state. In Section \ref{2} we give theoretical predictions for the experimental results including the imperfections of the setup: the losses and the multi-mode structure of twin beams. Section \ref{3} is devoted to the description of the experiment and the discussion of the results obtained for the bright squeezed vacuum state.  Section \ref{4} presents the measurement results obtained for classical (thermal and coherent) beams. We discuss the possible application of multi-photon HOM interference for quantum filtering in Section \ref{5}. We finish with the conclusion.


\section{Twin-beam bright squeezed vacuum}
\label{1}

A twin-beam bright squeezed vacuum state can be produced via high-gain parametric down-conversion (PDC), by strongly pumping a $\chi^{(2)}$ nonlinear crystal with appropriate phase matching. It is described by the state vector
\begin{equation}
\ket{\Psi_\text{in}}=\sum_{n=0}^\infty c_n \ket{n}_i\ket{n}_s,
\label{BSV}
\end{equation}
with the subscripts $s,i$ labeling the signal and idler beams, the coefficients $c_n=\tanh^n G/\cosh G$, and $G$ denoting the parametric gain. In our experiment, type-II phase matching is used and the two beams are orthogonally polarized. In the general case, they can be distinguishable in other parameters, such as wavevector direction or wavelength. In most experiments with bright squeezed vacuum, multimode twin beams are produced, with pairs of modes being independent. As a result, the total state vector is a product of state vectors (\ref{BSV}) for different mode pairs.

Clearly, the photon-number difference for each pair of modes is zero and does not fluctuate in the absence of losses. This effect is observed in experiment by measuring the noise reduction factor, namely the variance of the photon-number difference in the twin beams normalized to the mean photon-number sum~\cite{Lugiato,Bondani,Brida2,Iskhakov2009,Agafonov2010}. This measure is robust against multimode detection as both the variances and the mean values of photon numbers for independent modes sum up to give the photon-number variance and mean for the whole beam~\cite{Agafonov2010}.
\vspace{5mm}

\section{Interference}
\label{2}

To observe the interference of the two beams, the state is $45^\circ$ polarization rotated by a half-wave plate (HWP) and sent to a polarizing beam splitter (PBS). This transforms the orthogonally polarized input modes, described by $\hat{a}_s$ and $\hat{a}_i$, into the orthogonally polarized output modes $\hat{b}_1$ and $\hat{b}_2$ as follows:
\begin{align}
\label{eq:bs_trafo}
\begin{bmatrix}
\hat{b}_1^\dagger\\
\hat{b}_2^\dagger
\end{bmatrix}
=
\begin{bmatrix}
\sqrt{\tau}\ee^{\ii\varphi_\tau} & \sqrt{\varrho}\ee^{\ii\varphi_\varrho} \\
-\sqrt{\varrho}\ee^{-\ii\varphi_\varrho} & \sqrt{\tau}\ee^{-\ii\varphi_\tau}
\end{bmatrix}
\begin{bmatrix}
\hat{a}_s^\dagger\\
\hat{a}_i^\dagger
\end{bmatrix},
\end{align}
with $\tau$ and $\varrho$ being the transmissivity and reflectivity of the PBS, fulfilling $\tau+\varrho=1$. The quantities $\varphi_\tau$ and $\varphi_\varrho$ are phases associated with the PBS.
Following \cite{Campos}, we obtain the output state
\begin{align}
\label{eq:psi_out}
\ket{\Psi_\text{out}}&=\sum_{n=0}^\infty c_n\sum_{N=0}^{2n} R_N^{(n,n)}\ee^{\ii\varphi(N-n)}\ket{N,2n-N},
\end{align}
where $\varphi=\varphi_\tau+\varphi_\varrho$ and the coefficient $\bigl[R_N^{(n,n)}\bigr]^2$ gives the probability to obtain $N$ photons at output port 1 and $2n-N$ photons at output port 2 of the PBS when $n$ photons enter each of its input ports. It is defined as
\begin{subequations}
\label{eq:RNnn}
\begin{align}
R_N^{(n,n)}&=\sqrt{\frac{(2n-N)!}{N!}[\tau(1-\tau)]^{N-n}}\frac{(2N-2n)!}{(N-n)!}\nonumber\\
&\quad\times C_{2n-N}^{(N-n+1/2)}(2\tau-1),
\end{align}
for $N\geq n$ and
\begin{align}
R_N^{(n,n)}&=(-1)^{N-n}R_{2n-N}^{(n,n)},
\end{align}
\end{subequations}
for $N<n$ with $C_n^{(\alpha)}(x)$ being the Gegenbauer polynomials.

Let us consider the interference of the bright squeezed
vacuum state~(\ref{BSV}) on a PBS preceded by an ideal HWP, i.e.\ $\tau=0.5$. We are interested in the photon-number distribution at the output of the
BS. If we restrict ourselves only to events with $\sigma$
output photons in total, the photon-number distribution at any output port of the PBS will be the same, up to a constant factor, as for the case of two Fock states with photon numbers $\sigma/2$ simultaneously arriving at the input. The distribution is proportional to the discrete arcsine law,
\begin{align}
P(N|\sigma)=\begin{cases}
  |c_{\sigma/2}|^2/2^\sigma\binom{N}{N/2}\binom{\sigma-N}{(\sigma-N)/2},&\sigma\text{ even},\\
  0,&\sigma\text{ odd}.
\end{cases}
\end{align}
\begin{figure}[htb]
\centering
\includegraphics[width=0.8\textwidth]{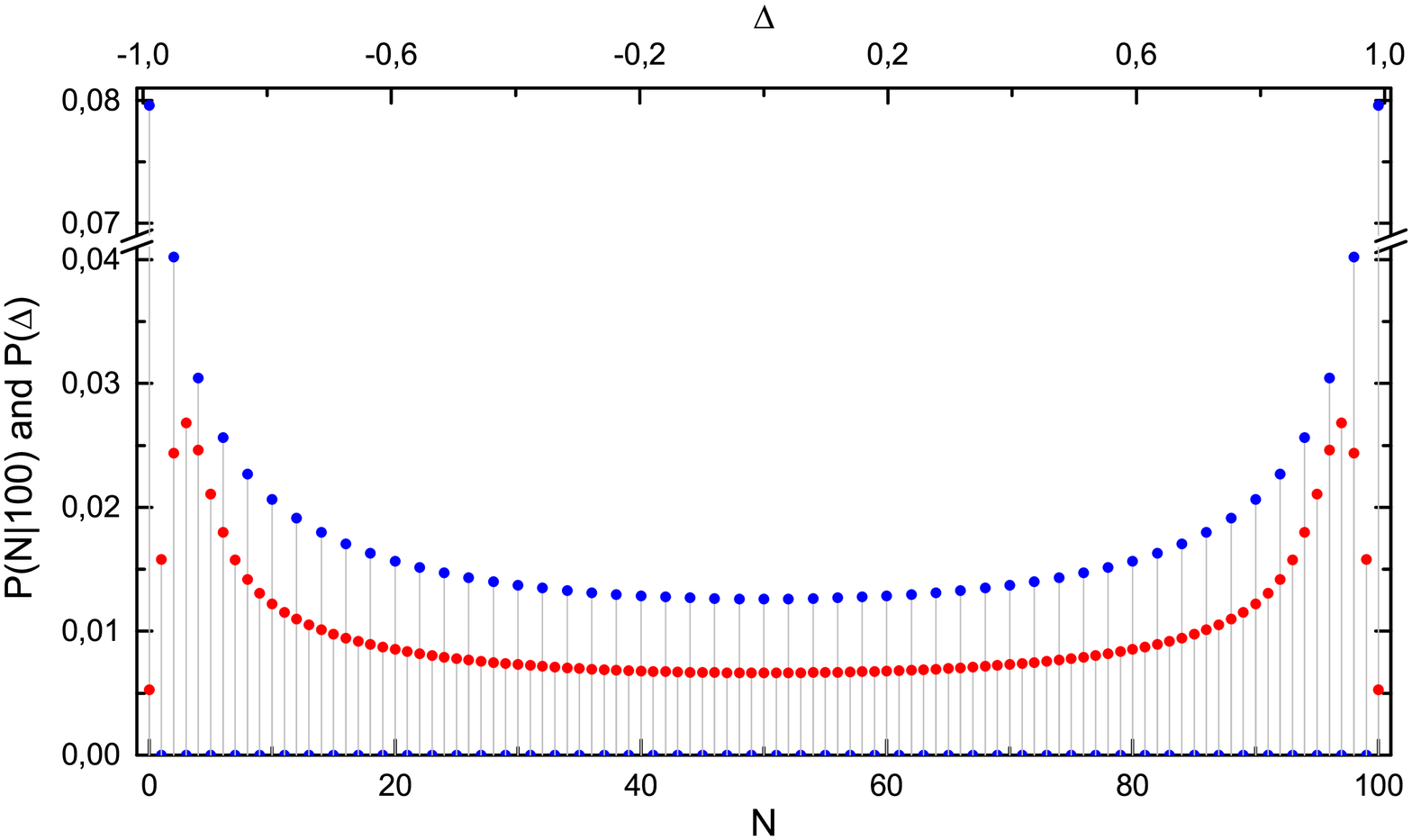}
\includegraphics[width=0.8\textwidth]{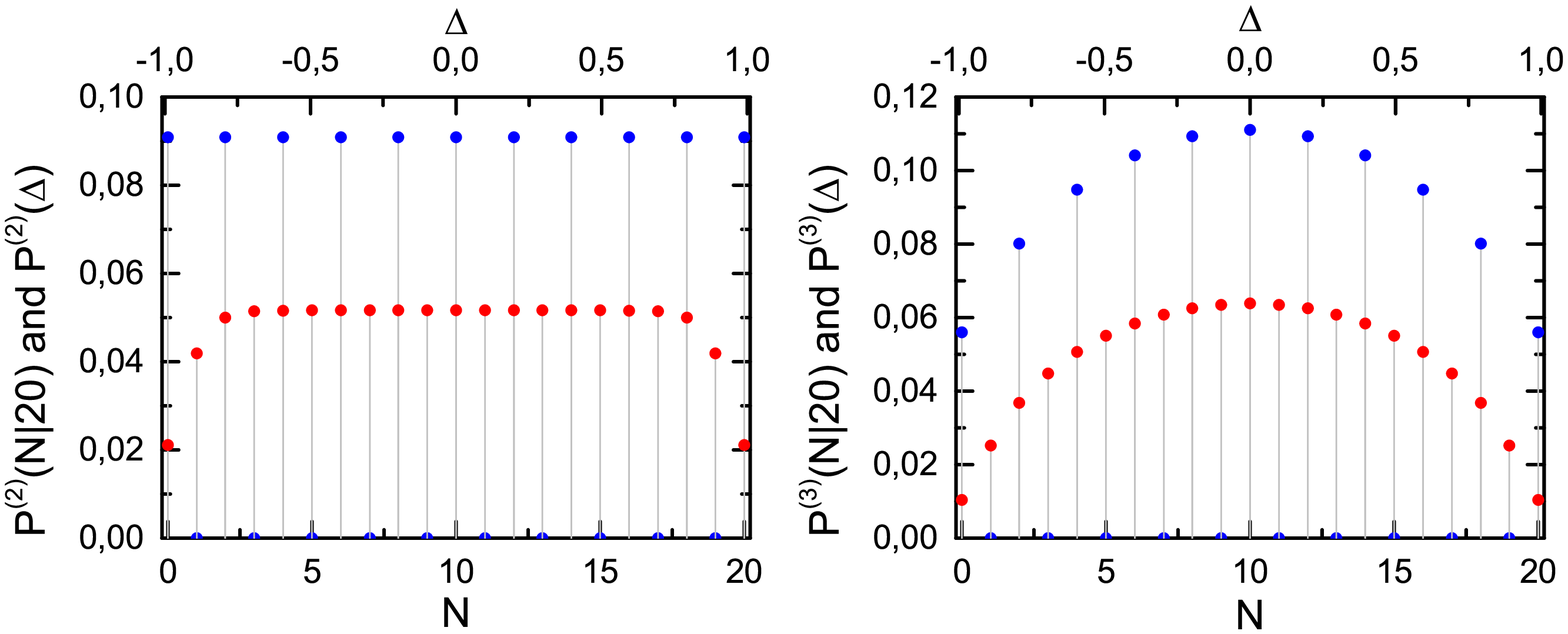}
\caption{Top: Photon-number distribution $P(N|\sigma)$ for $\sigma=100$ and its rescaled version $P(\Delta)$. Bottom: Photon-number distribution $P^{(m)}(N|\sigma)$ and its rescaled version $P^{(m)}(\Delta)$ for $\sigma=20$ and $m=2$ (left) as well as $m=3$ (right). Blue points show the lossless case ($\tau=0.5$ and $\eta=1$) and red points indicate the realistic case ($\tau=0.35$ and $\eta=0.05$).}
\label{fig.pn}
\end{figure}

Top panel of Fig.~\ref{fig.pn} shows this distribution for $\sigma=100$ (blue dots). The envelope manifests a typical U-shape. Notice that odd photon numbers do not occur and the maxima are at $N=0$ and $N=\sigma$. This type of distribution occurs, for instance, in the context of a classical one-dimensional
random walk~\cite{Campos}. It gives the probability that a one-dimensional classical random walker (a `drunkard') visited the initial point for the last time on step $N$ after $\sigma$ steps have been made. Note, however, that the position of the same walker after $\sigma$ steps is given by the binomial distribution, and exactly the same binomial law will describe the photon-number probability distribution for \textit{independent} (distinguishable) Fock states at the inputs of the beam splitter.

In experiment, it is more convenient to measure, instead of $P(N|\sigma)$, the probability distribution of the combination $\Delta=\tfrac{N_1-N_2}{N_1+N_2}$, where $N_{1,2}$ are photon numbers at the output ports 1,2 of the PBS. The shape of the probability distribution $P(\Delta)$ is a rescaled version of $P(N|\sigma)$, with the abscissae restricted to values between -1 and 1. At the same time, measurement of the distribution $P(\Delta)$ does not require postselection of the data and therefore can be performed with a larger sampling.

The state (\ref{BSV}) describes only a single mode in each of the twin beams. In the case of $m>1$ independent modes, the multi-mode distribution $P^{(m)}(N|\sigma)$ is given by a discrete convolution of the single-mode distributions $P(N|\sigma)$. Explicitly for two modes this means $P^{(2)}(N|\sigma)=\sum_{n=0}^N\sum_{s=0}^\sigma P(n|s)P(N-n|\sigma-s)$. At the bottom of Fig.~\ref{fig.pn}, photon-number distributions for two and three modes are shown, by blue dots. For two modes, the distribution is flat and has zeros at odd numbers. When considering three modes, we see a peaked distribution with the same odd/even structure.

In our current setup, this single-mode state can only be realized when accepting huge losses. To model the experimental results we also have to take into account that the PBS is non-perfect, i.e.\ $\tau\neq0.5$. In order to include losses we use the following positive operator valued measure (POVM) to calculate the joint-event probability $P_{m_1,m_2}$ of measuring $m_1$ photons in the output port 1 and $m_2$ photons in the output port 2 \cite{vogel_welsch}:
\begin{align}
\hat{P}_{m_1,m_2}=:\prod_{l=j}^2\frac{(\eta\hat{n}_j)^{m_j}}{m_j!}\exp[-\eta\hat{n}_j]:,
\end{align}
where the colons denote normal ordering, $\hat{n}_j$ is the photon number operator of the PBS output mode $j\in\{1,2\}$ and $\eta$ is the detection efficiency for each of these modes. The expectation value of this operator for the state from Eq.~(\ref{eq:psi_out}) determines the joint-event probability
\begin{align}
\label{eq:pm1m2}
P(m_1,m_2)&=\bra{\psi_\text{out}}\hat{P}_{m_1,m_2}\ket{\psi_\text{out}}\nonumber\\
&=\sum_{n=\lceil\frac{m_1+m_2}{2}\rceil}^\infty|c_n|^2\sum_{N=m_1}^{2n-m_2}\bigl[R_N^{(n,n)}\bigr]^2\eta^{m_1+m_2}\nonumber\\
&\quad\times(1-\eta)^{2n-m_1-m_2}\binom{N}{m_1}\binom{2n-N}{m_2}.
\end{align}

Finally, we obtain $P(N|\sigma)\equiv P(N,\sigma-N)$, for the realistic case of $\tau=0.35$ and $\eta=0.05$ (depicted in Fig. \ref{fig.pn} by red dots). Due to the losses the odd/even structure, present in the ideal case, vanishes. In the single-mode case the bias of the PBS causes the maxima of the distribution to move towards the center. Such behavior is more pronounced for higher $\sigma$. Moreover, the higher the bias of the beam splitter, the closer the maxima get to each other, ultimately converging to one central peak.

\section{Experiment}\label{3}

\begin{figure}[htb]
\centering
\includegraphics[width=0.6\textwidth]{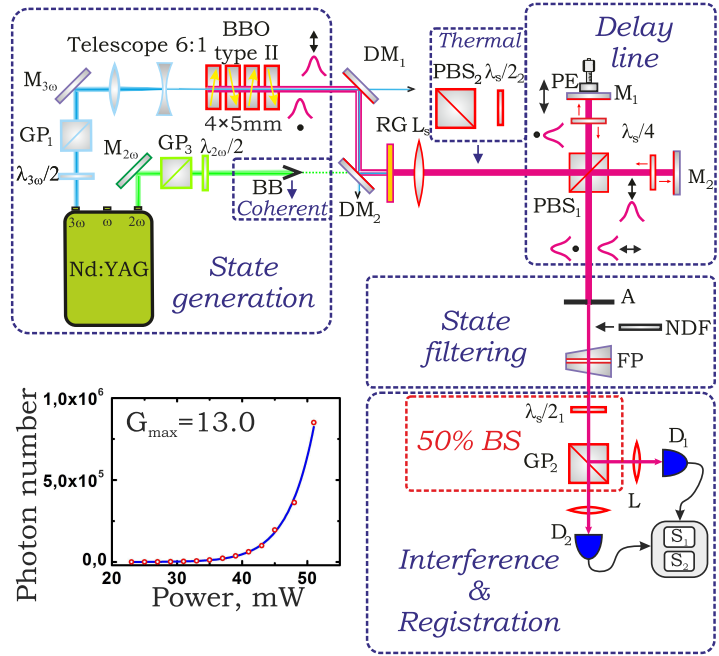}
\caption{Experimental setup. Inset: the dependence of the mean number of photons per PDC pulse on the pump power.}
\label{Setup}
\end{figure}

The experimental setup is presented in Fig.~\ref{Setup}. Twin beams were generated via high-gain PDC in four 5 mm BBO crystals with collinear frequency-degenerate type-II phase matching. The crystals had optic axes tilted in opposite directions to reduce the effect of spatial walk-off \cite{Slusher1987}. As a pump we used the third harmonic of a pulsed Nd:YAG laser at wavelength 355 nm. The pulse duration was $18$ ps, the pulse repetition rate $1$ kHz, and the energy per pulse up to $0.1$ mJ. The pump power was varied through the rotation of the half-wave plate ($\lambda_{3\omega}/2$) in front of a Glan prism (GP$_1$). In order to obtain higher PDC signal we focused the pump radiation using a telescopic system (Telescope 6:1), which consisted of a convex lens (F=50 cm) and a concave lens (F=-7.5 cm) at a distance of 42.5 cm. After the crystals the pump radiation was cut off by two dichroic mirrors (DM$_1$ and DM$_2$) with high transmission coefficient for the pump radiation and high reflection coefficient for the PDC radiation. The residual pump was absorbed by a long-pass filter RG630 (RG).

Due to the different group velocities of the ordinary and extraordinary beams, signal and idler pulses were always delayed with respect to each other, and we used a delay line to compensate for that. The signal and idler beams were split on a polarizing beam splitter (PBS$_1$) and then propagated along different paths to mirrors M$_1$ and M$_2$, respectively. The position of the mirror M$_1$ could be changed roughly using a micrometer and accurately using a piezoelectric actuator (PE). Polarization of the signal and idler beams was then rotated by $90^\circ$ by a double pass through quarter-wave plates ($\lambda_s/4$) rotated by $45^\circ$ degrees with respect to PBS$_1$. Therefore both beams exited through the same output port of PBS$_1$. The delay was controlled by observing the HOM interference effect~\cite{HOM}, as it is critical to the simultaneity of the signal and idler photons arrivals.

In order to obtain nearly single-mode signal and idler beams, spatial and spectral filtering was applied. The spatial filtering was performed by an aperture (A) with $0.8$ mm diameter placed in the focal plane of a convex lens with F=75 cm (L$_s$). The aperture selected an angle of $1$ mrad. The spectral filtering was performed by passing the radiation through a Fabry-Perot interferometer (FP) with a $250$ $\mu$m base. The size of the base was varied to achieve the PDC spectrum with only a single transmission maximum. To make the signal and idler beams indistinguishable, this maximum was shifted to the degenerate wavelength of the PDC radiation by slightly rotating the interferometer. The spectrum was monitored by a spectrometer HORIBA Jobin Yvon Micro HR (not shown in Fig.~\ref{Setup}). After the filtering, the width of the PDC spectrum was only $0.013$ nm.

A combination of a Glan prism (GP$_2$) and a HWP ($\lambda_s/2_1$) rotated by $22.5^\circ$ played the role of a 50\% beam splitter for the signal and idler beams. The reflected and transmitted beams were focused by a lens (L) to pin-diode based charge-integrating detectors (D$_1$ and D$_2$). The electronic signal from the detectors, representing the integral number of photons per pulse, was digitized by a fast analog-to-digital converter card \cite{Iskhakov2009}. The detectors always worked in the linear regime. The effective number of modes $m$ in the signal and idler beams was estimated from the $g^{(2)}\equiv\frac{\langle N_1N_2\rangle}{\langle N_1\rangle\langle N_2\rangle}$ measurement. It is well known that a single mode of the PDC signal or idler radiation has thermal statistics, i.e.\ $g^{(2)}_{0}=2$ \cite{Mollow1967,Tapster1998,Iskhakov2012}, and in the multi-mode case $g^{(2)}_{m}=1+\frac{g^{(2)}_{0}-1}{m}$~\cite{Ivanova2006}. Thus, from the measured $g^{(2)}_{m}$ we calculated the number of modes $m$ in the signal and idler beams. If the light was too bright for the detectors, the intensity was reduced using neutral density filters (NDF).

To find the parametric gain $G$, we measured the mean photon number per PDC pulse as a function of the pump power $P$. As the number of photons per mode in the PDC is $N_{mode}=\sinh^2(G)$, this dependence was approximated by the function \cite{Ivanova2006}
\begin{equation}
N=N_0\sinh^2\left(\kappa\sqrt{P}\right),
\label{PDC_sinh}
\end{equation}
with $N_0$ and $\kappa$ being the fitting parameters. Then the parametric gain was calculated as $G\equiv\kappa\sqrt{P}$.

At the maximum we obtained $G_{max}=13$ or $N_{mode}=4.9\times 10^{10}$ photons.
\begin{figure}[htb]
\centering
\includegraphics[width=0.4\textwidth]{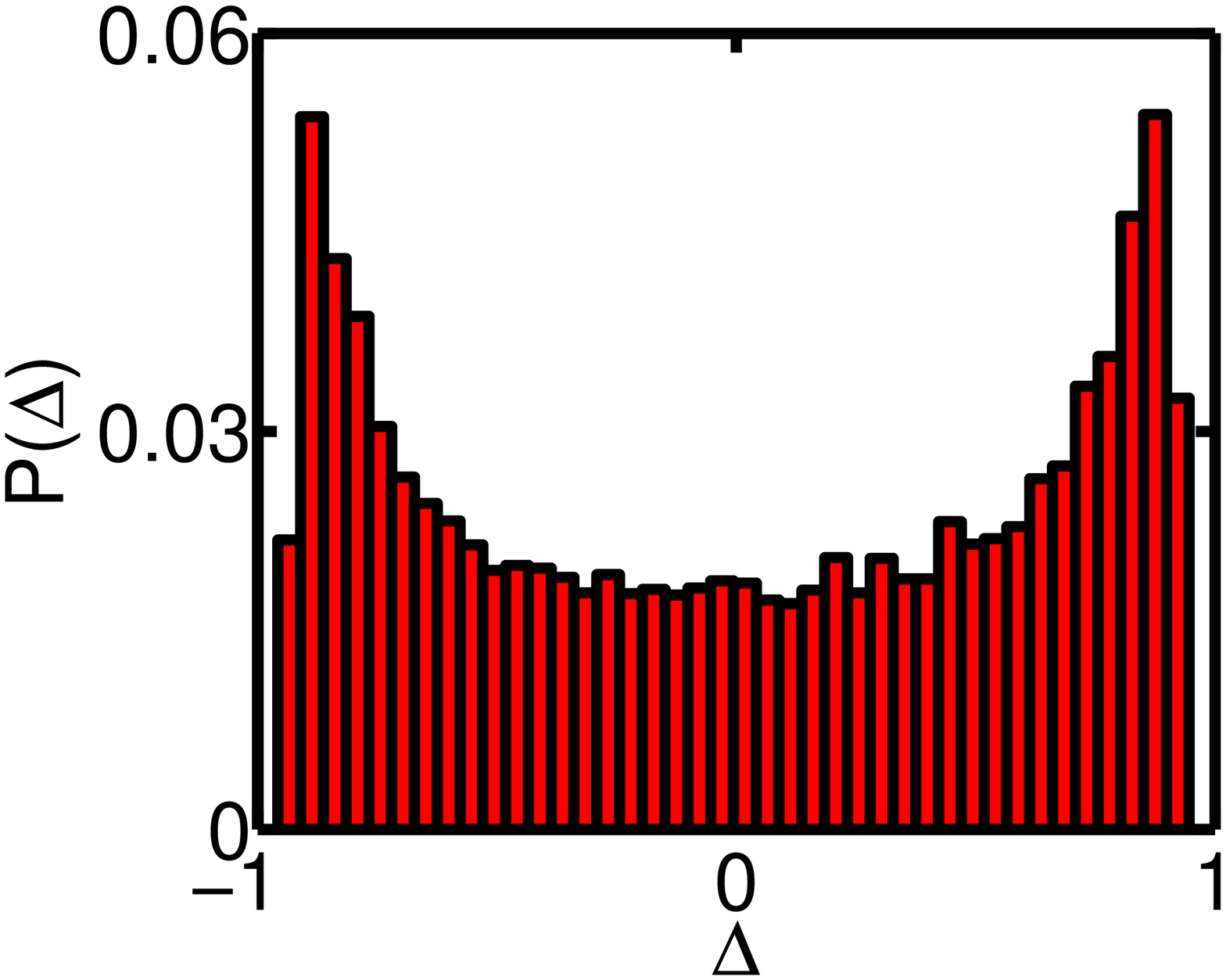}
\includegraphics[width=0.4\textwidth]{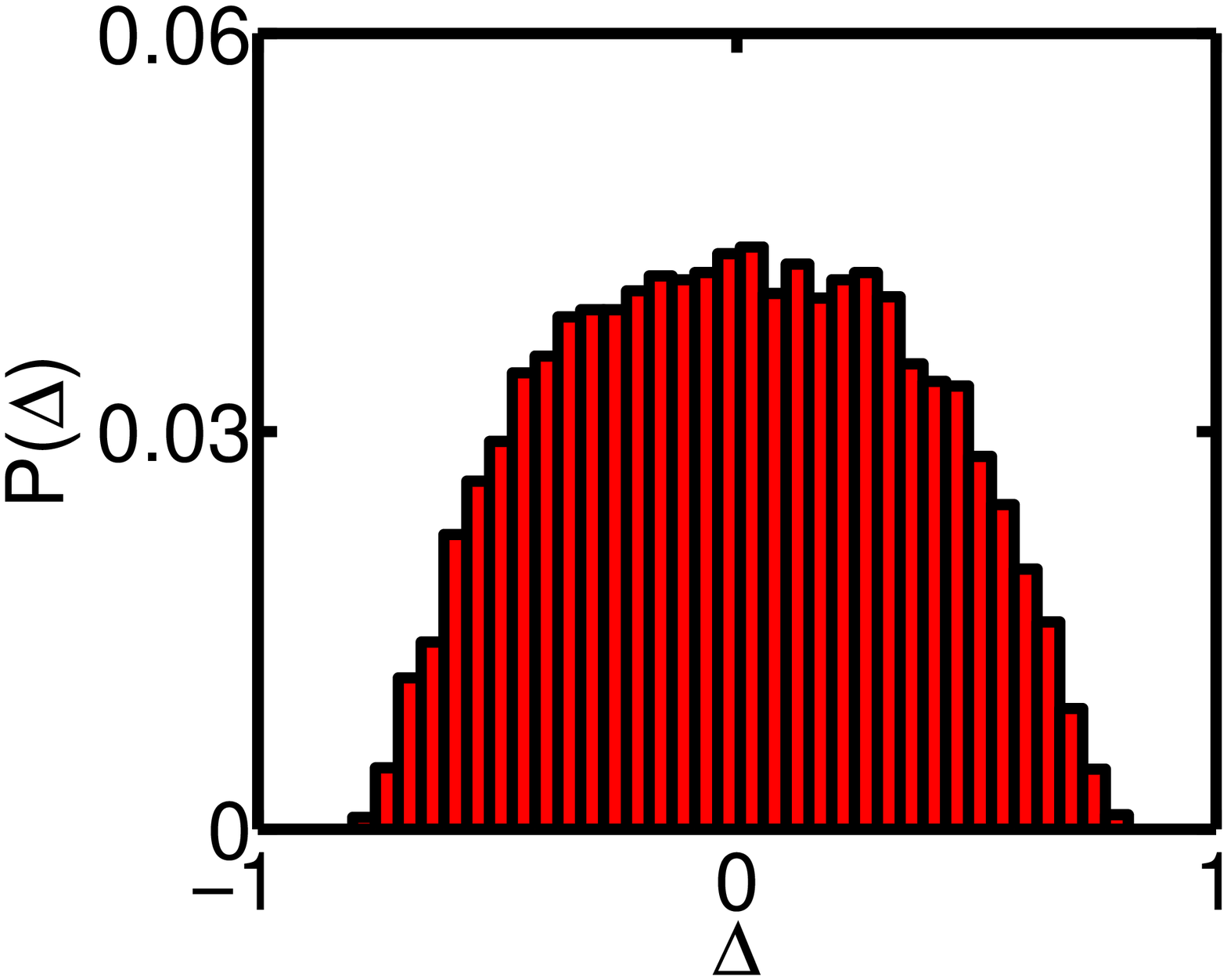}
\caption{Probability distributions $P(\Delta)$ obtained for the effective numbers of modes $m=1.2$ (left) and $m=3.4$ (right).}
\label{modes}
\end{figure}

In the presence of spectral filtering, nearly single-mode case ($m=1.2$) was obtained while without the filtering, few modes were observed ($m=3.4$). The results (Fig.~\ref{modes}) are in agreement with the theoretical predictions. For $m=1.2$ the probability distribution $P(\Delta)$ demonstrates the U-shape. This reflects the enhanced relative phase uncertainty between the twin beams at the beam splitter input. As it was shown theoretically, the distribution does not occupy all range from -1 to 1 because the beam splitter is unbalanced: the Glan prism always introduces 5\% losses into one polarization.
For $m=3.4$ we obtained a peaked distribution.
Both distributions have no odd/even structure, due to the losses and to the fact that the detectors do not have single-photon resolution.

Thus, both theoretical and experimental results show that the U-shape is not destroyed even in the presence of high losses in both beams (in experiment losses were higher than 99\% because of the spatial and spectral filtering) but disappears dramatically when several modes are present. The former happens because the losses do not reduce the large relative phase uncertainty of the twin beams. The latter occurs because in the multi-mode case, there is an ensemble of multiple U-shapes with different widths, due to the independent photon-number fluctuations in different modes, and the effect is averaged over this ensemble. It should be stressed that without photon-number fluctuations in different modes (e.g. for a mixture of independent Fock states) the U-shape would still be observed.

\begin{figure}[htb]
\centering
\hspace{0.065\textwidth}\textbf{a}\hspace{0.392\textwidth}\textbf{b}\\
\includegraphics[width=0.4\textwidth]{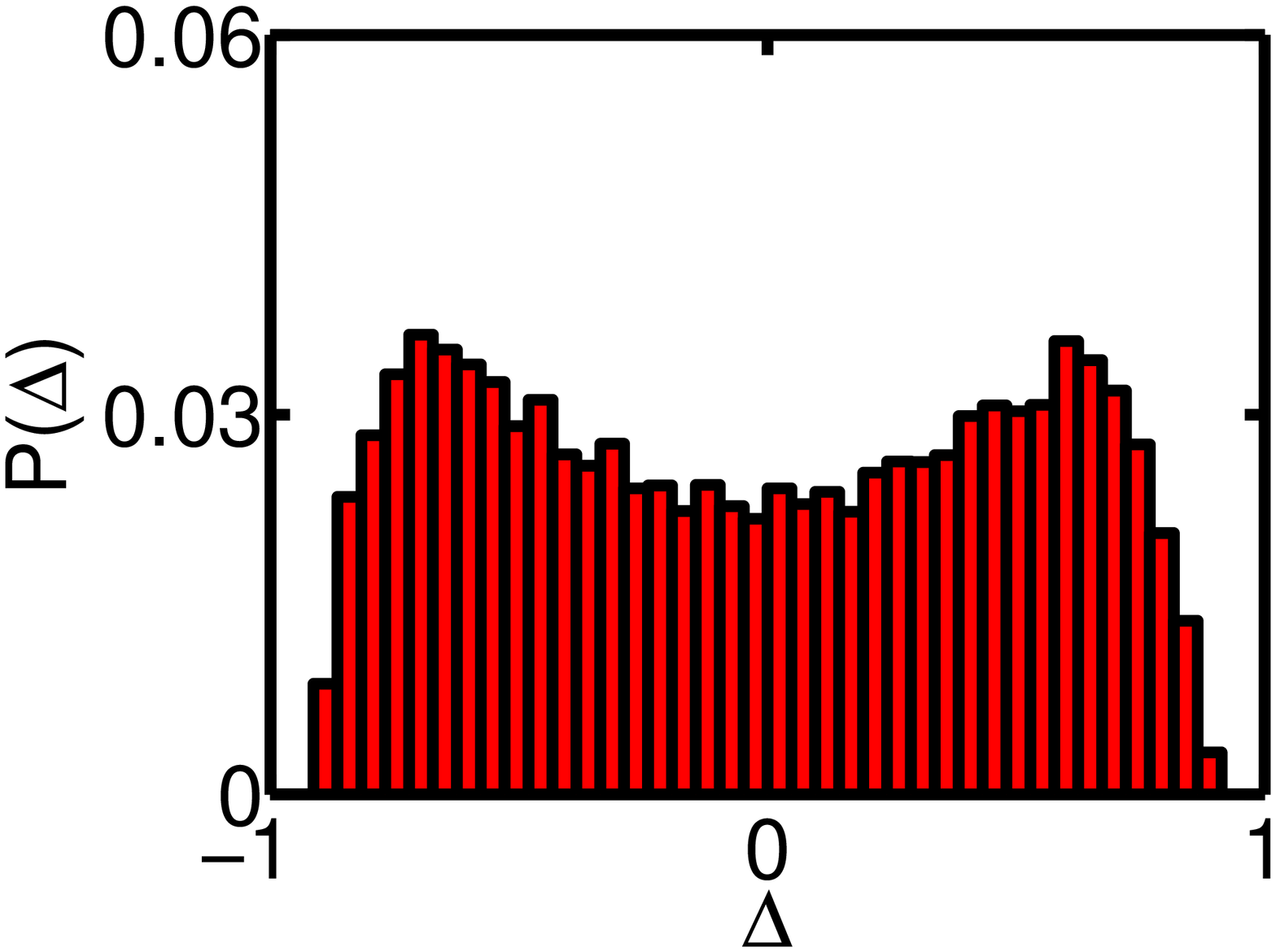}
\includegraphics[width=0.4\textwidth]{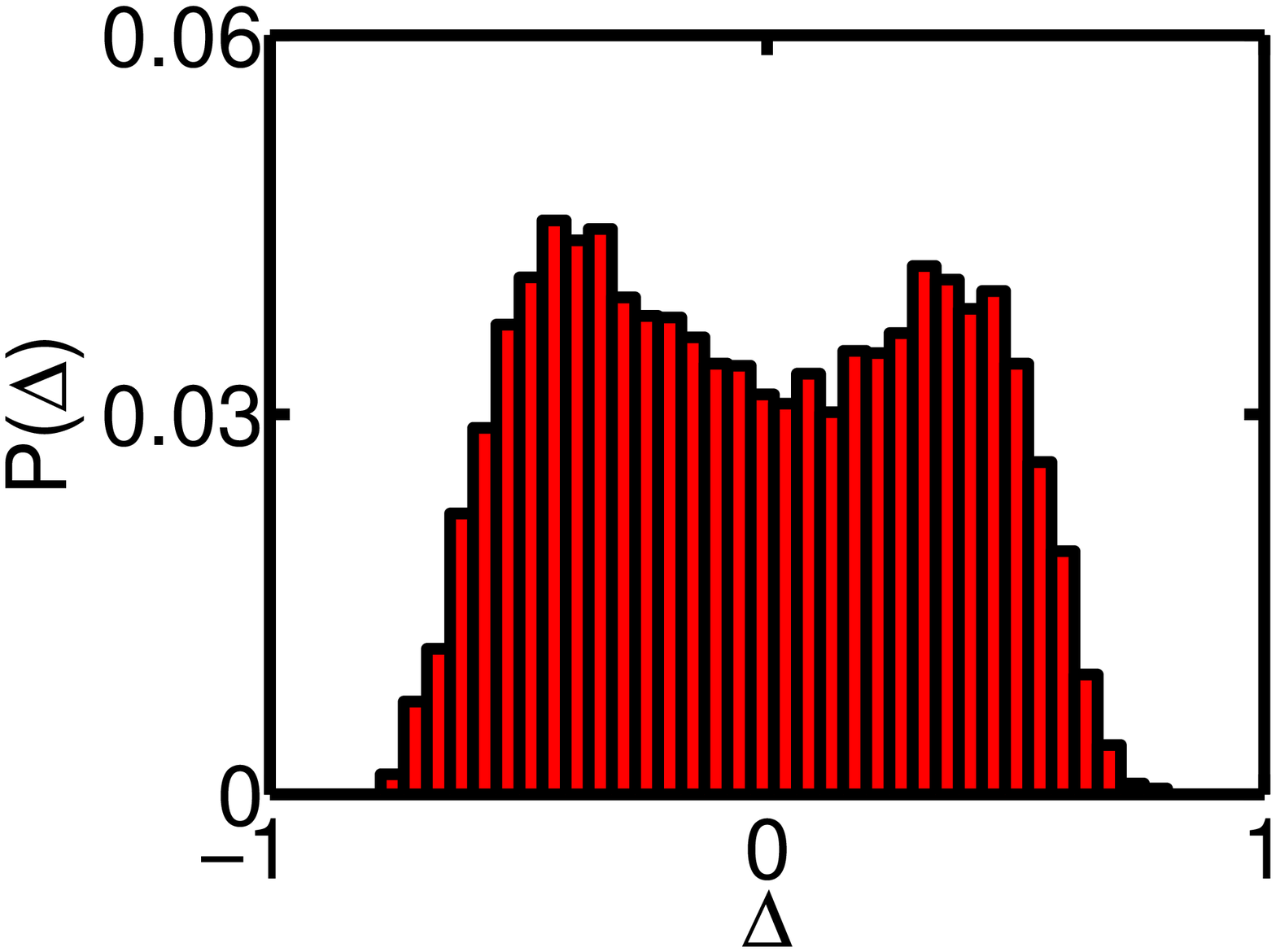}\\
\hspace{0.065\textwidth}\textbf{c}\hspace{0.392\textwidth}\textbf{d}\\
\includegraphics[width=0.4\textwidth]{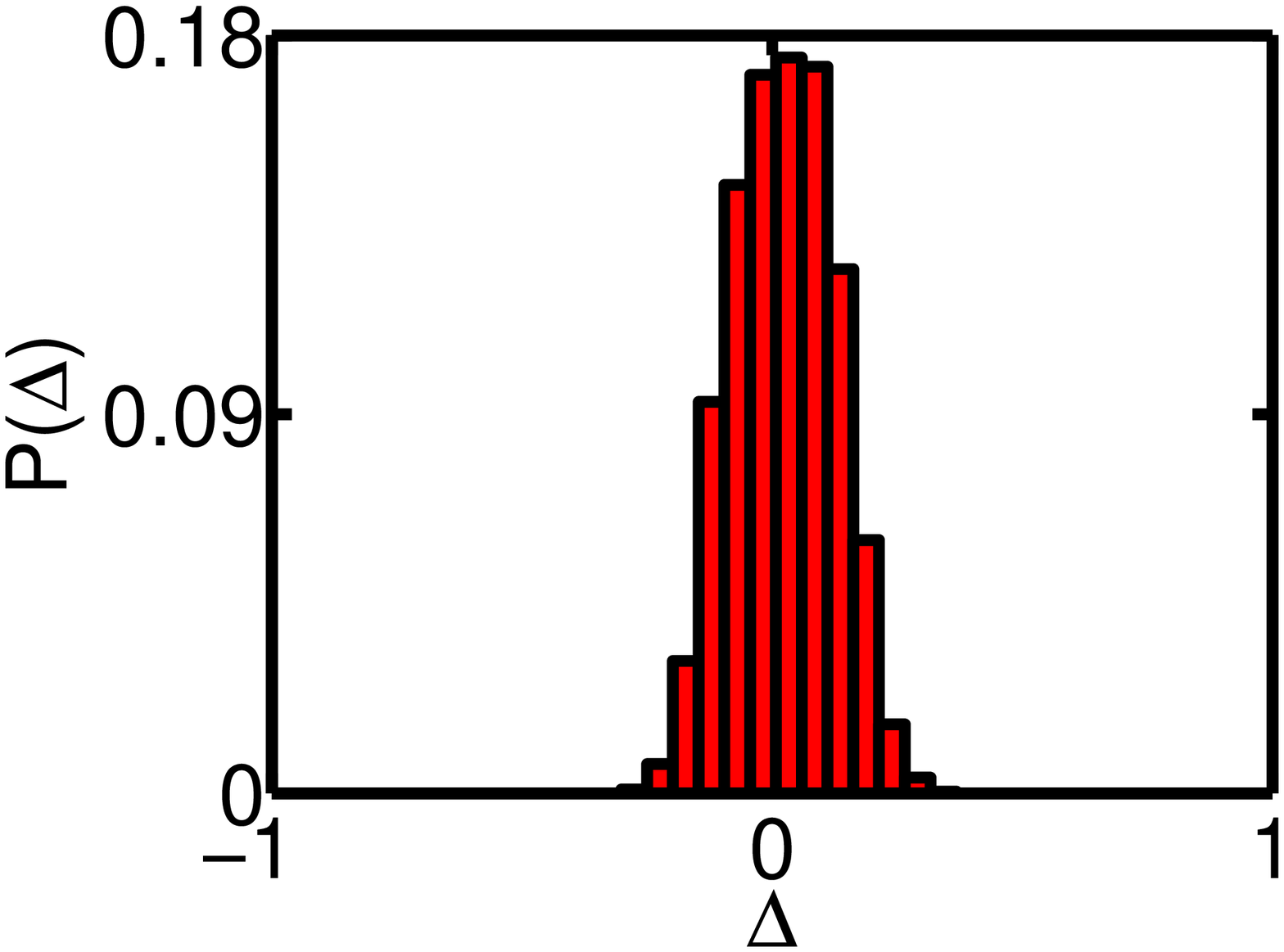}
\includegraphics[width=0.4\textwidth]{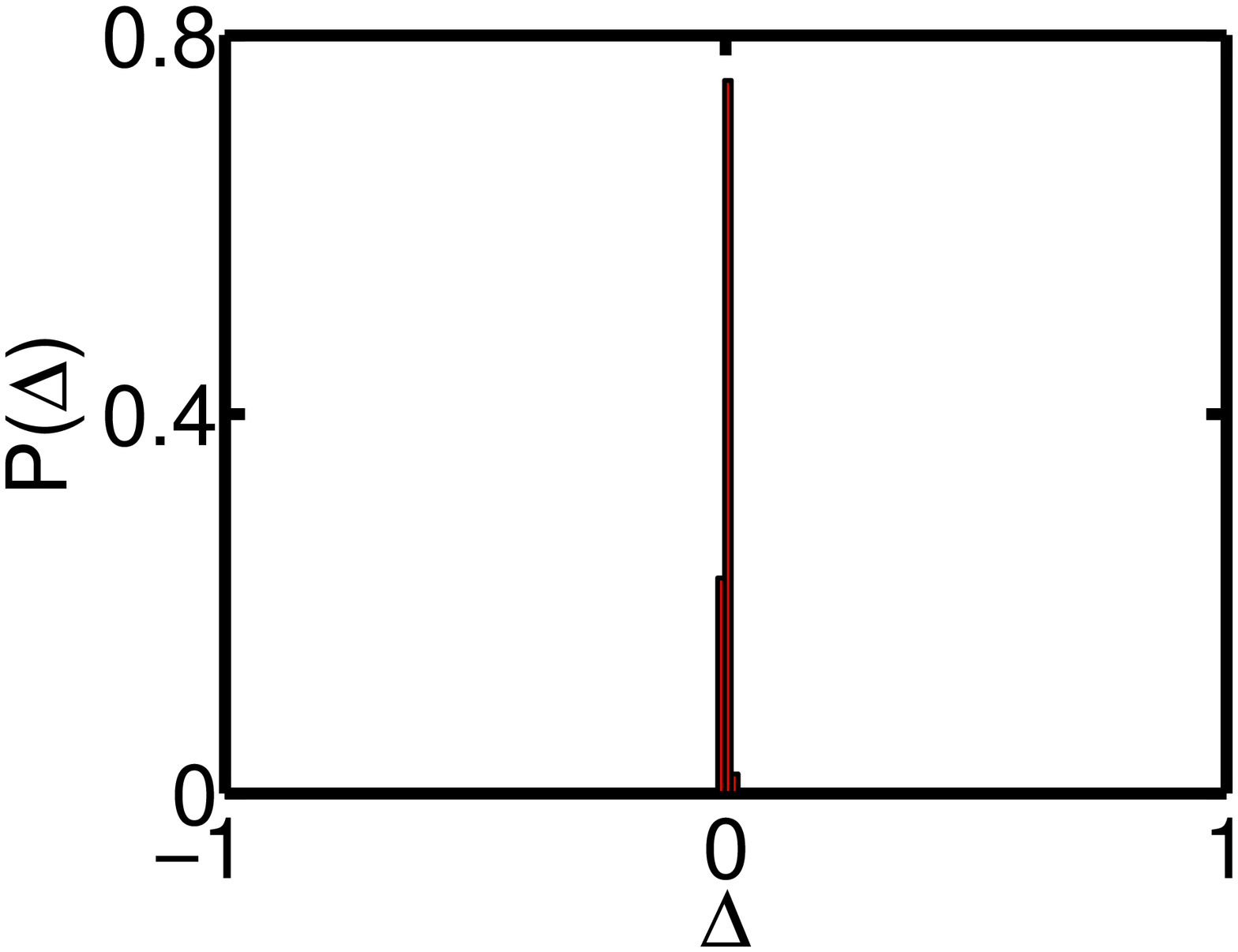}
\caption{Probability distributions $P(\Delta)$ obtained for the case of $m=1.2$ effective modes and different ratios of the mean photon numbers of the signal and idler beams: $1:3$ (a), $1:5$ (b), $1:17$ (c) and $1:\infty$ (d).}
\label{ratio}
\end{figure}


We also studied the interference of non-equally populated signal and idler beams (Fig.~\ref{ratio}). This case was obtained by introducing losses in the signal beam alone. It is noteworthy that the shape remains convex up to a considerable (5:1) ratio between the input photon numbers (Fig.~\ref{ratio}b), a feature mentioned in Ref.~\cite{Kasevich} in connection with the robustness of the phase resolution to the unbalance between the twin beams. The initial U-shape transforms then to a single peak (Fig.~\ref{ratio}c) and in the limit of one beam completely blocked, the distribution has a shot-noise limited width (Fig.~\ref{ratio}d). The same behavior was theoretically predicted for Fock states in Ref.~\cite{Campos}. Note that a single peak with the shot-noise limited width will also appear in the case of two independent Fock states at the beam splitter inputs.

\section{Classical beams}
\label{4}

The U-shape can be also observed for classical beams with the fluctuations in the relative phase increased artificially. This classical analogy was pointed out in Ref.~\cite{Campos}: whenever the phase of a classical oscillator is distributed uniformly, the quadrature will have the arcsine probability distribution, imitating the U-shape. This can be observed via homodyne detection of a classical beam whose phase is randomly modulated. Even simpler, one can split a classical beam in two, vary the phase of one of them, overlap both beams on a beam splitter and measure the distribution of the photon-number difference at the output. These very experiments we have performed with thermal and coherent sources. As thermal radiation, we used the signal PDC beam, filtered to contain $m=1.2$ effective modes. To separate the signal beam from the idler one we introduced another polarizing beam splitter (PBS$_2$), see Fig.~\ref{Setup}. The signal radiation was split by means of PBS$_1$ preceded by a half-wave plate ($\lambda_s/2_2$) at $22.5^\circ$. Further, the signal radiation interfered at the 50\% beam splitter. To randomize the phase, we applied a fast varying voltage to the piezoelectric actuator.

\begin{figure}[htb]
\centering
\includegraphics[width=0.4\textwidth]{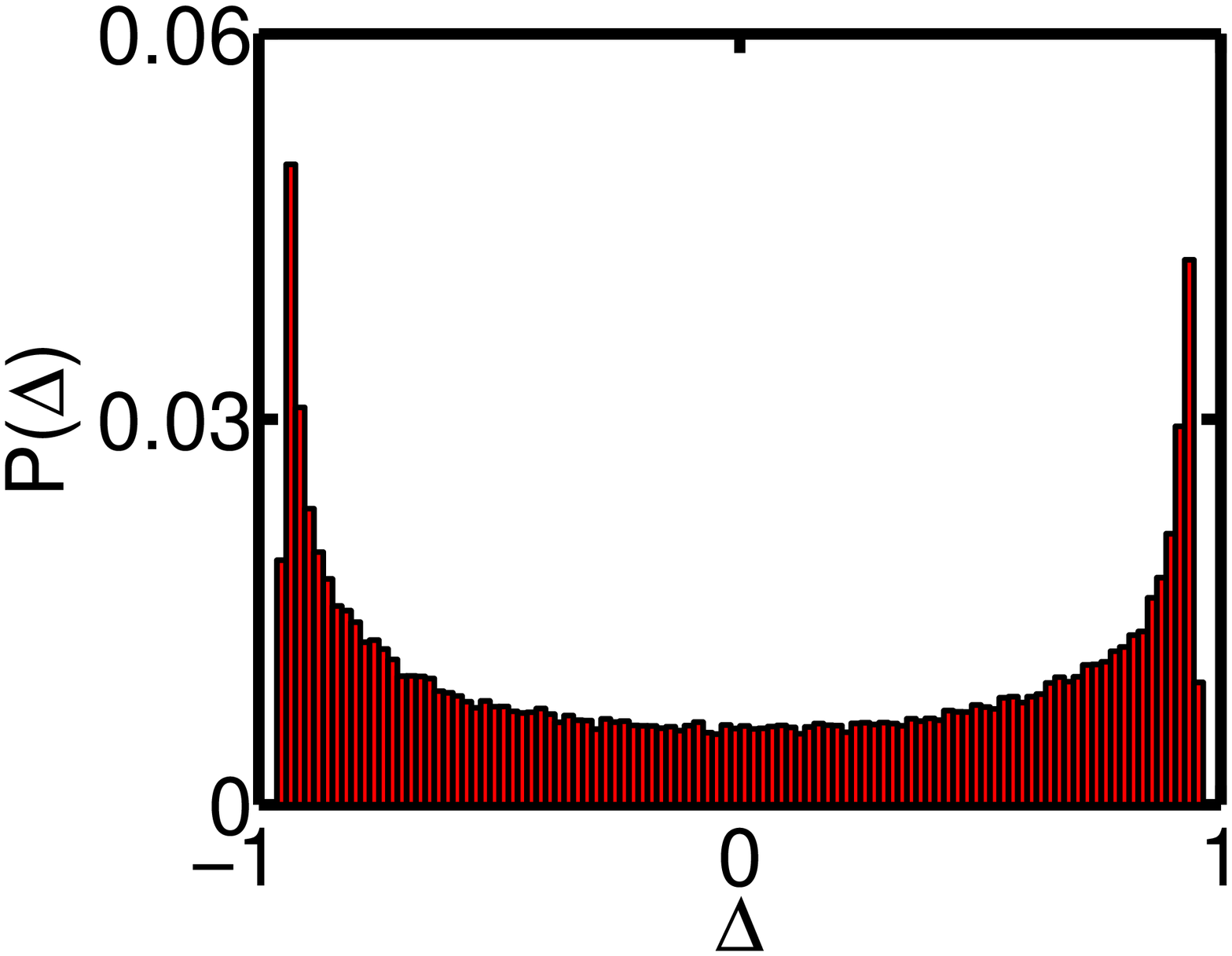}
\includegraphics[width=0.4\textwidth]{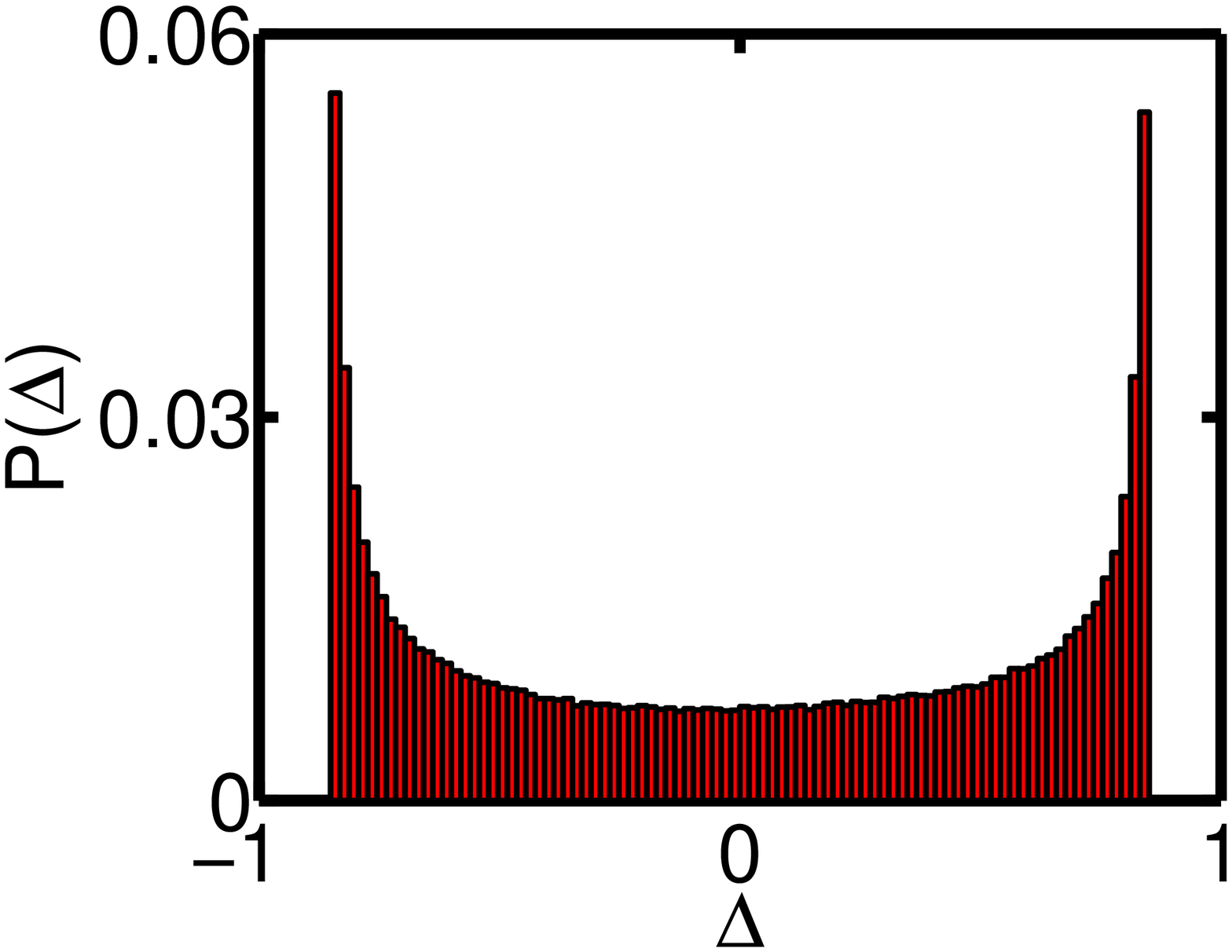}
\includegraphics[width=0.4\textwidth]{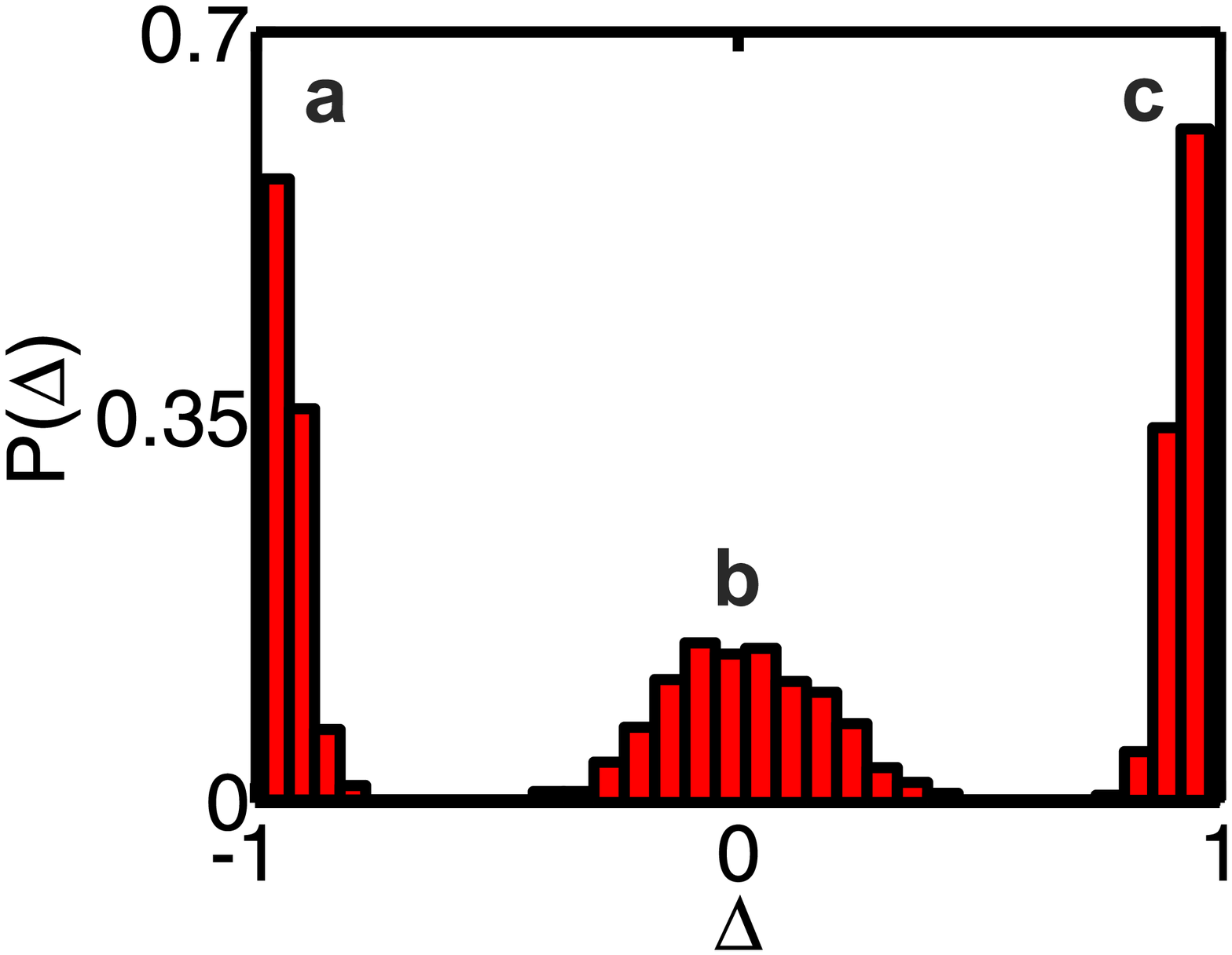}
\caption{The probability distribution $P(\Delta)$ for the thermal (upper left) and coherent (upper right) states with the randomized phase. Below: Three typical distributions $P(\Delta)$ for the thermal states for different phases (a,b,c).}
\label{Thermal&Coherent}
\end{figure}

The results are presented in Fig.~\ref{Thermal&Coherent}~(top left). The probability distribution $P(\Delta)$ demonstrates the U-shape for the randomized phase, and becomes peaked if the phase is fixed (bottom).

As a coherent beam, we used the second harmonic of the Nd:YAG laser (Nd:YAG 2$\omega$) at wavelength 532 nm (see Fig.~\ref{Setup}). As in the case of thermal light, a HWP ($\lambda_{2\omega}/2$) oriented at $22.5^\circ$ with respect to the PBS$_1$ was used to split the beam. The beam block (BB), the long-pass filter (RG), and the Fabry-Perot interferometer (FP) were removed in this case.

The probability distribution $P(\Delta)$ (top right panel of Fig.~\ref{Thermal&Coherent}) shows the same U-shape for the randomized phase, but is narrower than the one for the thermal case. This was caused by the non-ideal interference, due to the fact that the optical elements were not optimized for 532 nm.

It should be stressed that despite the broad photon-number distribution observed for classical light, phase super-resolution should be in no way expected here because the phase fluctuations have been increased artificially, by `mixing' the state. In contrast, for a pure twin-beam state the relative phase between the twin beams is fundamentally uncertain. This uncertainty is similar to the one of the Fock states, and it leads to a reduced uncertainty in the relative phase after the beam splitter. This is why bright twin beams are considered as candidates for beating the standard quantum limit for phase measurements~\cite{Holland,Giovannetti,Anisimov}. It is only due to losses and the absence of single-photon resolution that the Heisenberg limit seems to be impossible to achieve. This is a manifestation of the general problem of coarse-grained measurements~\cite{coarse}.

\section{Filtering macroscopic superpositions}
\label{5}

Apart of revealing the relative phase uncertainty of the twin beams, the observed effect can have an important application for quantum state engineering. The distributions shown in Figs.~\ref{modes} (left) and \ref{ratio} clearly show the following tendency: the more symmetric are the photon numbers at the input, the more asymmetric they are at the output, and vice versa. This suggests that the interference effect described above can be used for filtering out two-mode macroscopically populated states of light with either close or much different photon numbers, such as bright squeezed vacuum~\cite{Stobinska2012} or displaced path-entangled single photons~\cite{Sekatski2012}. At the same time, as pointed out in Ref.~\cite{filter}, such a filter would not provide information on the photon numbers at the two input ports of the beam splitter as the distributions at the two output ports are symmetric. It can be considered as a device measuring the modulus of photon-number difference but not the sign of this difference. 

That feature can be useful when working with macroscopic superpositions of Schr\"odinger-cat type as the ones considered in Ref.~\cite{DeMartini},
\begin{equation}
|\Phi\rangle=\sum_{i,j=0}^{\infty}\gamma_{ij}|2i+1,2j\rangle\mathrm{~~and~~}
|\Phi_\bot\rangle=\sum_{i,j=0}^{\infty}\gamma_{ij}|2j,2i+1\rangle,
\end{equation}
where notation $|k,l\rangle$ means a Fock state of $k$ photons in polarization mode $\phi$ and $l$ photons in polarization mode $\phi_\bot$ and $\gamma_{ij}$ are the amplitudes. Although the two states $|\Phi\rangle$ and $|\Phi_\bot\rangle$ are orthogonal, a coarse grained detector can not distinguish them very well because of a big effective overlap in the photon number distribution. The filter will enhance the components of the macroscopic states with $|i-j|\gg\delta$ for any given $\delta$ and thus reduce the effective overlap allowing real detectors to distinguish between the two states better. Nevertheless, superpositions of $|\Phi\rangle$ and $|\Phi_\bot\rangle$ will not be destroyed since both states produce the exactly the same response of the filter~\cite{filter,Buraczewski2012}. Therefore these superpositions can be employed as a macroscopic qubit in quantum protocols.

\section{Conclusion}
We have measured the probability distribution of photon-number difference at the outputs of a balanced beam splitter with bright twin beams fed to its inputs. The distribution has a typical convex shape with large standard deviation, reflecting the relative phase distribution for the beams at the beam splitter input. As this phase is Heisenberg conjugate to the photon-number difference, it has enhanced uncertainty. In an ideal (lossless) situation, this behaviour would provide the relative phase after the beam splitter defined better than allowed by the standard quantum limit. In a real experiment, however, such phase super-resolution would be reduced due to losses. The observation of the convex shape is only possible provided that a single mode is selected; this has been done in our experiment at the cost of more than 99\% losses introduced, which made impossible any attempts towards phase super-resolution. Additionally, we have shown the classical analogue of the effect for thermal and coherent beams with randomly modulated phase. This shows clearly that the U-shape in the photon number difference distribution alone is not an indicator for reduced uncertainty in the relative phase. Finally, we have discussed the application of this effect to the filtering of macroscopic superposition states.

\ack
We would like to thank Pavel Sekatski, Marek \.{Z}ukowski, and Farid Khalili for helpful discussions. We acknowledge partial financial support of the EU FP7 under grant agreement No. 308803 (project BRISQ2), the Russian Foundation for Basic Research, grants 11-02-01074 and 12-02-00965, and NATO SPS Project 984397 ``SfP-Secure Quantum Communications''. K.~Yu.~S. acknowledges support from the Dynasty Foundation.  M. S. was supported by the EU 7FP Marie Curie Career Integration Grant No.~322150 ``QCAT'', NCN grant No.~2012/04/M/ST2/00789, FNP Homing Plus project No.~HOMING PLUS/2012-5/12 and MNiSW co-financed international project No.~2586/7.PR/2012/2.

\end{document}